\input{aipcheck}

\documentclass[
    ,final            
  ]
  {aipproc}
\newcommand{\be}{\begin{equation}}
\newcommand{\ee}{\end{equation}}
\newcommand{\bea}{\begin{eqnarray}}
\newcommand{\eea}{\end{eqnarray}}

\newcommand{\bn}{\bar\nu}

\newcommand{\nnu}{\nonumber\\}

\newcommand{\sqt}{\sqrt{3}}

\newcommand{\oot}{\overline {126}}

\newcommand{\boot}{${\bf{\oot}}$ }

\layoutstyle{6x9}

\begin{document}

\title{Emergence of the NMSGUT}\footnote{Plenary talk at IWTHEP, Roorkee, India, 15-20 March,
2007.} \classification{12.10g,12.10 Dm, 12.10 Kt,12.15 Ff}
\keywords {Minimal Supersymmetric SO(10) GUT}

\author{Charanjit S. Aulakh }{  address=  {Dept. of Physics, Panjab University
 Chandigarh, India 160014  }}

\begin{abstract}
We  trace  the  emergence  of the `` New Minimal'' supersymmetric
SO(10) GUT (NMSGUT) out of the debris created  by  our
demonstration that the MSGUT is falsified by the data.  The NMSGUT
is based on ${\bf{210\oplus 10\oplus 120\oplus 126\oplus
{\overline {126}} }}$ Higgs system. It has only spontaneous CP
violation and Type I seesaw. With only 24 real superpotential
parameters  it is the simplest model capable of accommodating the
known  18 parameter fermion mass data set and yet has enough
freedom to accommodate the still unknown Leptonic CP violation and
neutrino mass scale parameters. Our focus is on the two most
salient features uncovered by our analysis: the domination of the
\boot Yukawa couplings by those of the ${\bf{10,120}}$ (required
for evasion of the no-go that trapped the MSGUT) and the
inescapable raising of the Baryon violation scales(and thus
suppression of proton decay) decreed by a proper inclusion of the
threshold effects associated with the calculated superheavy
spectra. These two structural features are shown to be
complementary.

\end{abstract}

\maketitle


\section{ Introduction }
 Following
  the discovery of neutrino mass  Supersymmetric SO(10)
theories, particularly ones based on the ${\bf{210 \oplus
126\oplus {\overline {126}} }}$ Higgs system, have received a
great deal of attention\cite{msgutlong,TypeI,TypeII}. SO(10)
unification has multiple virtues that seem to pre-destine it for
canonicity: a SM family {\it{together with a conjugate neutrino}}
fits in one chiral $\mathbf{16}$-plet of SO(10). Using the  \boot
Higgs  both the Type I and Type II seesaw mechanisms find a
natural implementation. The $\mathbf{210}$-plet Higgs together
with the $\mathbf{126\oplus {\overline {126}}}$
breaks\cite{aulmoh} SO(10) to the MSSM  but preserves R-parity as
a part of its gauge symmetry, so that the theory presents its
stable LSP as   dark matter. The  simple superpotential
allows\cite{abmsv} an explicit and complete solution of the
spontaneous symmetry breaking in terms of a single complex control
parameter ($x$)  and complete calculation of the superheavy
spectrum. The parameter $x$ satisfies a simple cubic equation
dependent linearly  on  a parameter ratio ($\xi$): so the map
$x\rightarrow\xi$ is $1-to-1$. The `SU(5)
conspiracy'\cite{rparso10} enforced by RG flows calculated from
the computed spectra -as is required in the Susy
case\cite{aulmoh,surviv}- restricts one  to the single step
breaking case.  From the spectra  detailed information on the
parameter space of the theory can be deduced at an unprecedented
level of control and refinement thus pinning down the theory so
that it is rendered falsifiable. Unexpectedly this yields an
explanation for the suppression of proton decay\cite{newmsgut}.

The other aspect  concerns  the generic features of fermion mass
spectra in SO(10).   Fermion masses and mixings (22 parameters in
all) extrapolated to the superheavy scale using a grand desert
assumption are the primary data that any Susy GUT must
accommodate.  An early attempt to implement a complete fermion
data fit using only the $\mathbf{10 +\oot}$
representations\cite{babmoh} set the rules for much of the later
work. After initial difficulties the increasingly precise
observation of neutrino masses  stimulated and facilitated the
  successful `generic' fits of all fermion data using
both Type I and Type II seesaw mechanisms\cite{TypeI,TypeII}.
Thus, till 2005, the Babu-Mohapatra program seemed   successful.
 However it was just assumed that the   overall scale and
relative strength of   Type I  versus  Type II seesaw masses were
realizable  in  Susy GUTs.   Our MSGUT survey\cite{gmblm} revealed
 serious difficulties in obtaining Type II over
Type I dominance as well as in obtaining large enough Type I
neutrino masses.
 Using a convenient parametrization   in
terms of the  single ``fast''   parameter ($x$) which controls
MSGUT  ssb we gave\cite{blmdm}  a complete proof   of the failure
of the Seesaw mechanism in the context of the MSGUT. The nature of
the obstruction uncovered by us   led us to suggest\cite{blmdm} a
  natural scenario that   dealt with the problem by extending
the MSGUT with a $\mathbf{120}$ FM Higgs representation. In this
  scenario\cite{blmdm,core} the $\mathbf{120}$ -plet
 and the $ \mathbf{10}$ -plet   fit the dominant
charged fermion masses. Small $\mathbf{\oot} $-plet couplings give
appreciable contributions only to light charged fermion masses
{\textit{and}} enhance  Type I seesaw masses  to viable values
since Type one seesaw masses are \emph{inversely} proportional to
these couplings. In what follows we give details of a new and
simpler Minimal Supersymmetric GUT candidate that has emerged from
the implementation of our suggestion. We will then   summarize the
challenges that this so called New MSGUT(NMSGUT)  faces in its
attempt to reach the status of  a falsifiable and predictive
theory.

\section{  Seesaw Failure   in the MSGUT}

  The Type I and Type II
   neutrino masses  in the MSGUT are\cite{blmdm}  :
\bea M_{\nu}^I &=& (1.70 \times 10^{-3} eV) ~ { F_{I}}~
\hat{n}~{\sin \beta}\nnu
 M_{\nu}^{II} &=& (1.70 \times 10^{-3} eV) ~{ F_{II}}
 ~\hat{f}~{\sin \beta}\nnu
    \hat{n}&=& ({\hat h} -3 {\hat f}) {\hat f}^{-1}
    (  {\hat h} -3 {\hat f})\nonumber
     \eea
where  $\hat{h},\hat{f}$ are proportional to
 the $\mathbf{10,\oot}$ Yukawa couplings
 to fermions, $\beta$ is
 the MSSM Higgs  vev ratio parameter, and
   $F_I(x),F_{II}(x)$ are  specified
  in the MSGUT, for explicit forms  see\cite{blmdm}.

In typical BM-Type II fits\cite{TypeII}  the maximum eigenvalue$
\hat{f}_{max}$     $ \sim 10^{-2}$ while   $ \hat{h}_{max} $ is
$\sim 1 $ so $\hat{n}_{max} \sim 10^2 $.
 Thus $R={F_I/F_{II}}  \leq 10^{-3}$ in order that
  the pure BM-Type II    not be overwhelmed by
   the BM-Type I values.  Such R values
    are  un-achievable\cite{gmblm,blmdm}
    in the MSGUT parameter space  (while preserving
  baryon stability, perturbativity etc).    Furthermore
in the BM-Type I fits,     typically  $ \hat{n}_{max} \sim 5
\hat{f}_{max} \sim .5  $ for the maximal eigenvalues of $
\hat{n}$. Thus values of $F_I\sim 100$ are required in order to
reach realistic values of seesaw masses for the heaviest neutrino.
We demonstrated\cite{gmblm,blmdm} that such values are not even
nearly  achievable   over the  complex $x $ plane except where
they also violate some aspect of successful unification. The same
argument also applies to mixed Type I and Type II\cite{TypeI}.
Later another group independently confirmed our
results\cite{bertnew}.

\section{The new $\mathbf{10-120-\oot}$ scenario }

    To resolve the difficulty with the overall neutrino mass scale
 we proposed\cite{blmdm,core}
    that  $\mathbf{\oot}$ yukawa couplings be reduced
    (well below the level where they are important for 2-3 generation
    masses)  and introduced a $\mathbf{120}$-plet to do the work of
    charged fermion mass fitting previously accomplished by ${\bf{\oot}}$
  couplings.
  The small $\mathbf{\oot}$ yukawa coupling would boost the value
  of the Type I seesaw masses by increasing $\hat n$  (and
  suppress Type II seesaw masses).
 Note   that an extension of the MSGUT by
  $\mathbf{120}$-plet, with only spontaneous CP violation but with  Type II dominant seesaw mechanism
   had been considered previously  \cite{dattmim}.
  However the assumption of the viability of the Type II generic fits used there has
  already been explained to be invalid.  Thus  our  scenario with
  Type I dominant and  a   characteristic and distinct  pattern of yukawa
  couplings (which has proven successful and viable) must
  be considered as a qualitatively  different
  theory. As we shall see this  also applies to the mechanisms of
  lowering  proton decay rates advocated in  the two
  scenarios.

  The Dirac masses in such GUTs are then
generically given by\cite{blmdm,newmsgut}
\bea   m^u &=&  \textrm{v}( {\hat h} + {\hat f} + {\hat g} )\nnu
  m_{\nu}&=&
 \textrm{v} ({\hat h} -3 {\hat f}  + r_5' {\hat{g}})
 \nnu
  m^d &=& { \textrm{v} (r_1} {\hat h} + { r_2} {\hat f}  +
r_6 {\hat g}) \\
    m^l &=&{ \textrm{v}( r_1} {\hat h} - 3 {  r_2} {\hat f} + r_7{\hat
   g})\nnu
  {\hat g} &=&2ig {\sqrt{\frac{2}{3}}}(\alpha_6 + i\sqt
\alpha_5)\sin\beta \quad;\quad \hat h = 2 {\sqrt{2}} h \alpha_1
\sin\beta  \quad;\quad\hat f = -4 {\sqrt{\frac{2}{3}}} i
f\alpha_2\sin\beta      \label{120mdir}\nonumber\eea

 The right handed neutrino mass is $M_{\bn}  = \hat{f}
\hat{{\bar{\sigma}}}$ and the Type I seesaw formula is \bea
M_{\nu}^{I} =   v r_4 \hat{n} \quad ;\quad \hat{n}&=& ({\hat h} -3
\hat{f} - r_5' {\hat g}) {\hat f}^{-1}
    (  {\hat h} -3 \hat{f }+ r_5'{\hat g})\eea
 where $\hat{{\bar{\sigma}}}     =  {i\bar{\sigma}\sqrt{3}}/
    {\alpha_2 \sin \beta} $, $\bar{\sigma} $
is the GUT scale  vev of the $\oot$ while  $\alpha_i,\bar\alpha_i
$ refer to fractions of the MSSM doublets contributed by various
doublets present in the GUT Higgs representations
\cite{abmsv,bmsv,ag2,gmblm,blmdm}. See \cite{newmsgut} for the
form of the coefficients $r_i$ in terms of the
$\alpha_i,\bar\alpha_i $ and the expression for the
$\alpha_i,\bar\alpha_i $ in terms of the parameters of the NMSGUT.

As regards spontaneous CP violation one finds that it requires the
parameters in the superpotential to be real and the six
independent phases\cite{grimus2,grimus3} in the generic fermion
mass matrices must thus arise from the various doublet vevs or, in
our language, where $<h_i>\rightarrow \alpha_i \textrm{v}_u,
<\bar h_i>\rightarrow \bar\alpha_i \textrm{v}_d$, from the
complexity of of the coefficients $\alpha_i,\bar\alpha_i $. This
can be shown\cite{newmsgut} to require that the fast control
parameter `$x$' of the MSGUT and NMSGUT superheavy spectra is
itself complex while the superpotential parameter ratio $\xi$ that
it determines (via the complex solution branches of the
fundamental cubic equation satisfied by $x$) is real. The values
of the other (slow) superpotential parameters are restricted to a
narrow range $\sim 1$ by unification constraints. Thus one can
actually scan\cite{ag2,gmblm,blmdm,newmsgut} the complete
behaviour of the theory as function of a single relevant real
parameter $\xi$. Subsequent semi-analytic  work by
us\cite{core,msgreb}  but more effectively  the purely numerical
analysis in\cite{grimus2,grimus3}  lent decisive support to our
fitting scenario.

In \cite{grimus2}   CP violation is spontaneous but free
parameters are reduced to 21 from 25  by fixing
$f_{12}=f_{23}=g_{13}=0$ by imposing a $Z_2$ symmetry. The
``downhill-simplex" nonlinear fitting method is used to minimize a
$\chi^2$ fitting function constructed from the values of 18
observables (9 charged fermion masses, 4 CKM parameters, 2
neutrino oscillation mass splittings, 3 PMNS angles) and the
experimental uncertainities thereof. The eigenvalues found are

\bea \hat h &:& 0.58~~\quad;~~-0.0213~~\quad;~~0.000375 \nnu \hat
f &:& 0.105  ~~;~~ 0.021~~;~~ 0.00033\nnu r_5'\hat g&:& \pm
 0.65~~;~~0\nnu \hat n &:&  31.19 ~~\quad;~~ 5.996 ~~\quad;~~  1.045  \eea

Similarly in \cite{grimus3} the same authors achieved even better
fits with fewer(18) parameters by fixing the 6 phases(6 parameters
less) so that CP was violated but without imposing any $Z_2$
symmetry(3 parameters more) on the Yukawas. They then obtained a
fit to the 18 known fermion mass parameters with the eigenvalues

 \bea \hat h &:&  0.473~~\quad;~~ 0.00037~~\quad;~~ 8.15\times 10^{-5}
\nnu \hat f &:& 2.45\times 10^{-3} ~~;~~1.49\times
10^{-3}~~;~~5.8\times 10^{-5}\nnu   r_5' \hat g&:& \pm  0.108
~~;~~0\nnu \hat n &:&  220.035 ~~\quad;~~ 39.867 ~~\quad;~~ 7.974
\eea Another similar fit but with inverted neutrino mass heirarchy
was also found. Clearly both fits illustrate how $f$ is reduced
and $\hat n$ enhanced sufficiently to allow accommodation in the
NMSGUT.
 In addition the fits fix the 1 Dirac and 2 Majorana phases
 of the leptonic mixing matrix and the overall neutrino mass scale.
  Since, however, these are not known
 at present it seems that the  parameter freedom fixed arbitrarily
 by these authors for numerical and illustrative purposes may
 finally be called upon to shoulder the task of also fitting the
 experimental values of these phases. In the NMSGUT\cite{newmsgut}
 there are 24  real superpotential parameters
 which are   reduced to 23 by
the fine tuning condition that keeps the MSSM doublets light  but
raised by two to count the electroweak vev and the susy parameter
$\tan \beta$ which are not fixed within the Susy GUT. The fermion
mass  data(12 masses + CKM phase + 3 CKM angles + 3 PMNS angles +
3 PMNS phases)  consists of 22 parameters at most. Thus somewhat
 disappointingly, even after reserving 4 parameters for the remaining mass data
 there  is still a surplus of 3 parameters left out of the total
 of 25 parameters. The excess of
three parameters  most likely implies that additional data such as
that from proton decay   will be required before the NMSGUT
becomes falsifiable or   its parameters determined. The situation
without a CP preserving Lagrangian is of course much(15 real
parameters more) worse.

\section{ Proton Decay scales in the NMSGUT  }

In \cite{ag2,gmblm,blmdm,newmsgut} we developed the analysis of RG
constraints on the NMSGUT following the approach of\cite{hall} in
which $M_X$ is taken to be the mass of the lightest  gauge
multiplet which mediates proton decay(and not the point where the
3 MSSM gauge couplings cross).  In this section we summarize
information  on the $x$ values allowed by imposing plausible
`realistic'  constraints on the magnitudes of the threshold
corrections to the  gauge couplings. We pay particular attention
to the scenario\cite{bs} where $M_X$ and with it all dangerous
$d=5$ proton decay mediating  Higgs triplet (of which there are
three distinct types $t[3,1,\pm {\frac{2}{3}}],P[3,3,\pm
{\frac{2}{3}}]$ and $K[3,1,\pm {\frac{8}{3}}]$ in this
theory\cite{newmsgut}) masses are pushed above $10^{16} GeV$.
     Quite reasonably, we  demand
\bea
|\Delta_G|&\equiv& |\Delta  (\alpha_G^{-1}(M_X))| \leq 10 \nonumber \\
2 \geq  \Delta_X &\equiv &\Delta (Log_{10}{M_X}) \geq - 1\nonumber \\
|\Delta_{W}|&\equiv &  |\Delta(sin^2\theta_W (M_S))|< .02
\label{criteria} \eea

to implement perturbative gauge dominance, $10\%$ uncertainity in
$\sin^2\theta_w(M_S)$ and $d=6$ proton decay mediation
suppression.

We find threshold corrections \cite{ag2,gmblm,blmdm,newmsgut}

\bea
 \Delta^{(th)}(Log_{10}{M_X})  &=& .0217 +.0167 \sum_{M'} (5 {{\bar b}'}_1
 +3{{\bar b}'}_2 -8
  {{\bar b}'}_3) Log_{10}{{M'}\over  {M_X }} \label{Deltasw}\nnu
\Delta^{(th)} (sin^2\theta_W (M_S)) &=&
   .00004 -.00024 \sum_{M'}(4 {{\bar b}'}_1 -9.6 {{\bar b}'}_2 +5.6
  {{\bar b}'}_3) Log_{10}{{M'}\over  {M_X }}
  \label{Deltath}\nnu
\Delta^{(th)} (\alpha_G^{-1}(M_X)) &=& .1565 +  .01832 \sum_{M'}(5
{{\bar b}'}_1 + 3 {{\bar b}'}_2 + 12 {{\bar b}'}_3)
Log_{10}{{M'}\over {M_X }}
  \eea
Where ${\bar b'}_i = 16\pi^2 b_i'  $ are   1-loop $\beta$ function
coefficients ($ \beta_i=b_i g_i^3 $)
 for multiplets with  mass $M'$.
  These corrections, together with the two loop gauge corrections,
  modify  the one loop values corresponding to the successful gauge
unification of the MSSM, see \cite{ag2,gmblm,blmdm} for details.

  The parameter $\xi= \lambda M/ \eta m$ is the only
  numerical parameter that  enters into the cubic
   eqn.   that determines the parameter $x$
   in terms of which all the  superheavy vevs are given.
    {\it{ It is thus  the most crucial  determinant of
     the mass spectrum }}.

     The rest of the coupling parameters divide into
     ``diagonal''($\lambda,\eta,\rho$) and ``non-diagonal''
     ($\gamma,\bar{\gamma},\zeta,\bar{\zeta }, k$)  couplings
     with the latter exerting a very minor influence
     on the unification parameters.  We have therefore
fixed the non diagonal parameters at representative values $\sim
1$ throughout.

A crucial point\cite{gmblm} is that the threshold corrections
depend only on ratios of masses and are independent of the overall
scale parameter which we choose to be the mass parameter $m$ of
the $210$-plet. Since $M_X=10^{16.25 +\Delta_X} GeV$ it follows
that $m\sim 10^{16.25 + \Delta_X }$. It is thus clear that this
factor will enter every superheavy mass so that they must all rise
or fall in tandem with $M_X$ i.e exponentially with $\Delta_X$.

 As an example of the regions of parameter space allowed by the
 unification constraints we present Figs. 1 and Fig. 2 from \cite{newmsgut}. It is clear that the
 bulk of the allowed parameter space has the mass of the lightest
 baryon violating gauge bosons, \emph{as well as all other superheavy
 masses including Higgs and Higgsino triplets that mediate $d=5$
 proton decay},   raised by a factor of 10 or more. This can be
 seen even more clearly in the case of the CP preserving
 superpotential which requires that we use the complex solution
 for $x$ and a real value for the parameter ratio $\xi$. The
 corresponding plot of $\Delta_X$ versus $\xi$ is given in Fig. 3
 for the region where there is a sensible variation of $\xi$. The
 constraint on $\Delta_W$ restricts one to $|\xi| <11$ while that on
 $\Delta_G$ excludes a narrow region around $\xi=-5$. As a result
 the values of all superheavy mass scales are raised well above
 the one loop unification scale of $10^{16.25} GeV$. We present a representative
 set of allowed values in Table 1. It is clear that the MSSM
 running in the grand desert will be practically unaffected and it
 is the modification of the relation between the one loop unification mass and
 the mass of the lightest baryon number violating gauge boson
 (and  thereby all other superheavy masses) that is responsible for
 the elevation of the dangerous superheavy masses.
 The upshot of the  mass scales raised in tandem is that both $d=6$
and $d=5$ baryon violation will be strongly suppressed over most
of the viable  NMSGUT parameter space.Thus we see that the  NMSGUT
  deals  with the difficulty regarding $d=5$
   mediated baryon decay in a very natural and unforced way.
    This may be contrasted with the mechanisms for suppression of
    proton decay in \cite{dattmim}.
    No fine tuning of  Yukawas or artificially arranged cancellation or any introduction of a plethora of
 uncontrollable  non-renormalizable terms is required at all. The
 suppression is generic in the viable regions of the parameter space and
 is practically inescapable  on     complex branches  of the
 ssb solution.

\begin{figure}
  \includegraphics[height=.3\textheight]{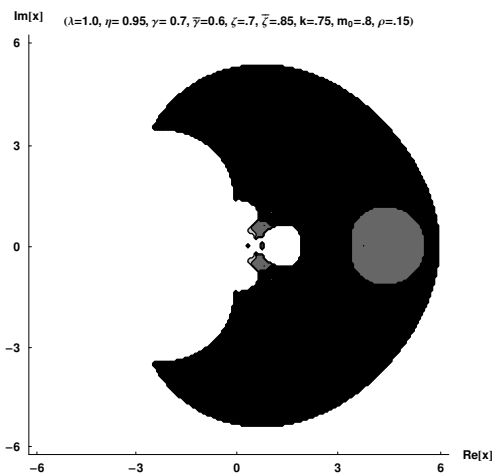}
  \caption{NMSGUT critical
behaviour: $\rho=\rho_c=.15$. Regions of the x-plane compatible
  with the unification constraints (\ref{criteria})
  are shaded. The darkest regions have $2\geq \Delta_X
  > 1 $ (corresponding to $M_X> 10^{17.25} GeV$), the next darkest $1\geq \Delta_X > 0$ the lightest shade
$0\geq \Delta_X > -1$   and the white regions are  disallowed.  }
\end{figure}

\begin{figure}
  \includegraphics[height=.3\textheight]{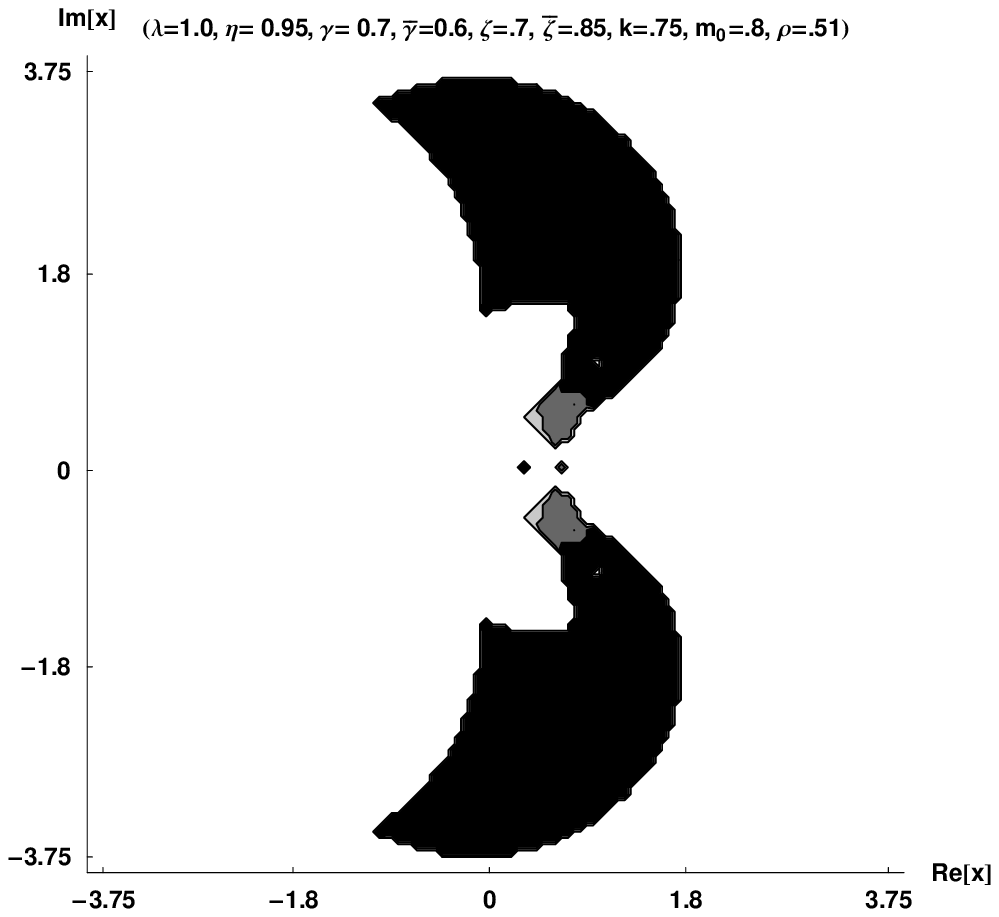}
  \caption{NMSGUT critical behaviour : $\rho=.51 $ i.e larger than
  $\rho_c=.15$ : Regions of the x-plane compatible
  with the unification constraints (\ref{criteria})  are shaded. The darkest regions have $2\geq \Delta_X
  > 1 $ (corresponding to $M_X> 10^{17.25} GeV$), the next darkest $1\geq \Delta_X > 0$ the lightest shade
$0\geq \Delta_X > -1$   and the white regions are  disallowed. }
\end{figure}

\begin{figure}
  \includegraphics[height=.3\textheight]{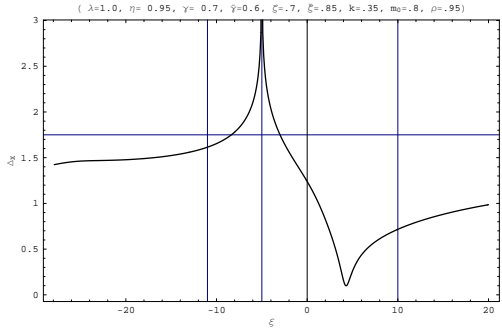}
  \caption{Plot of $\Delta_X$ against $\xi$ on the CP violating
 solution branch $x_+(\xi)$ at representative allowed values of the diagonal parameters. }
\end{figure}

\begin{table}

\begin{tabular}{|l|l|}
\hline
 {\rm Field }[SU(3),SU(2),Y] &\hspace{10mm} Masses ( Units\,\,  of $10^{18}$ Gev) \\
\hline A[1,1,4]  & 1.19 \\
B[6,2,{5/ 3}] & 0.49   \\
 C [8,2,1]  & {1.13, 0.91, 0.27} \\
 D[3,2,{7/ 3}]  &{2.03, 1.74, 0.47}\\
E[3,2,{1/ 3}] & {2.12,\, 1.56,\, 1.01,\, 1.01,\, 0.78,\, 0.55}   \\
F[1,1,2]& {2.16,\, 0.41,\, 0.41,\, 0.14}  \\
G[1,1,0]& 1.19, 0.85, 0.39, 0.33, 0.33, 0.06   \\
h[1,2,1]& {3.35, 2.46, 2.06, 1.34, 0.36}   \\
I[3,1,{10/ 3}]& 1.46   \\
J[3,1,{4/ 3}]& {1.99, 1.22, 0.61, 0.61, 0.22} \\
K[3,1, {8/ 3}]& 1.23, 0.49\\
L[6,1,{2/ 3}] &{1.45, 0.12} \\
M[6,1,{8/ 3}]  & 1.44 \\
N[6,1, {4/ 3}]  & 1.52 \\
O[1,3, 2]  & 3.86  \\
P[3,3, {2/ 3}] & 3.0, 0.21 \\
Q[8,3,0]& 1.29 \\
R[8,1, 0] & {2.39, 0.59} \\
S[1,3,0] & 2.64 \\
t[3,1,{2/ 3}]&{3.47, 2.26, 1.44, 1.03, 0.73, 0.18, 0.12} \\
U[3,3,{4/ 3}]  &  2.18\\
V[1,2,3] & 1.69  \\
W[6,3,{2/ 3}]  & 2.05\\
X[3,2,{5/ 3}]&0.98, 0.98, 0.6\\
Y[6,2, {1/ 3}]  & 0.7\\
Z[8,1,2] &2.37  \\
\hline \hline
 Parameter Values & $\lambda=1.0,\, \eta=.95,\, \gamma=.7,\, \bar{\gamma}=.6,\, \zeta=.7$\\
    & $\bar{\zeta}=.85,\, k=.35,\, m_0=.8,\, \rho=.95, \,m/\lambda = 5.28 \times 10^{16}Gev  $\\
  Min [\,Mass\,] ( Units\,\,  of $10^{18}$ Gev) & 0.06(G[1,1,0]), 0.12(\,L[6,1,2/3]\, , t[3,1,2/3]\,)  \\
   Max[\,Mass\,]( Units\,\,  of $ 10^{18}$ Gev) &   3.86(O[1,3,2]) \\
  $Threshold  Corrections $  & $\Delta_X = 1.739 ,\,\, \Delta_W=-0.0149,\,\, \Delta_G =0.4994 $\\
    &  $ \alpha_{G}=0.038$\\
Gauge Masses  ( Units\,\,  of  $10^{18}$ Gev)&
$m_{\lambda_E}=1.01,\, m_{\lambda_F}=0.41,\,
m_{\lambda_G}=0.67,$\\
  &\,$m_{\lambda_J}=0.61,\, m_{\lambda_X}=0.97 $ \\
 \hline
\end{tabular}
\label{table II} \caption{Example of Masses and couplings
favouring high $M_X$, $\xi=-8.5$ on the $x=x_+(\xi)$ branch for
real couplings  }$$
\end{table}

As a corollary to the raising of the scale of baryon violation,
however, the \boot vev responsible for the right handed neutrino
mass is also raised significantly so that compatibility with
Leptogenesis constraints also actually favours the small \boot
Yukawas that were already indicated by our fermion fitting
scenario. This is another instance of the interplay between
neutrino masses and baryon violation in SO(10). Another difficulty
alleviated by the raised mass scales is that the Landau pole in
the SO(10) gauge coupling that occurs once the superheavy
thresholds have intervened in the gauge evolution is pushed closer
to the Planck scale. This ties in well with our scenario\cite{tas}
that envisions a (calculable) UV
  condensation of coset gauginos  in the supersymmetric GUT
    which drives the breaking of the GUT
symmetry\cite{tas}.  Moreover since the cutoff of the perturbative
theory is about an order of magnitude less than  the scale  of UV
condensation and the Planck scale - which coincide, it is natural
to surmise that the gauge strong coupling dynamics induces gravity
characterized by $M_P\sim\Lambda_X$. It is worth remarking that
such a scenario naturally overcomes all the fundamental
objections\cite{david} that led to an abandonment of the induced
gravity scenarios of the 1980s.

To conclude : minimal SO(10) supersymmetric GUTs enjoyed a
 falsification which led to a  simpler   version compatible with all data.
 Minimality inspired investigations of Susy SO(10)
GUTs have thus  entered a kind of limbo. The vision\cite{babmoh}
of   predictive SO(10) Susy GUTs seems to have been an
un-realizable dream.    To break out the
  difficult  issues  of    fine tuned scale separation  and of
supersymmetry breaking must be faced: perhaps by tackling the
  fascinating issue of dynamical symmetry breaking due
to the inevitable asymptotic strength of the NMSGUT\cite{tas}.



\begin{theacknowledgments}
   It is pleasure to acknowledge discussions and collaboration
   with Sumit Garg, Borut Bajc, Alejandra Melfo and Goran
   Senjanovic. This work was supported by   grant No SR/S2/HEP-11/2005
from the Dept. of Science and Technology of the Govt. of India.
\end{theacknowledgments}


\end{document}